\documentstyle[12pt]{article}
\newcommand{\be}{ \begin{eqnarray}}
\newcommand{\ee}{\end{eqnarray}}
\newcommand{\beno}{ \begin{eqnarray*}}
\newcommand{ \eeno}{\end{eqnarray*}}
\newcommand{\raf}[1]{(\ref{#1})}
\begin{document}
\setcounter{page}{0}
\bibliographystyle{try}
\hspace{11cm}
{\large SUNY-NTG 94-3}
\vspace{.7cm}
\begin{center}
\ \\
{\large \bf
On the temperature dependence of  correlation functions in the space like
direction in hot QCD}
\vspace{2cm}
\ \\
{\large Volker Koch}
\ \\
{\it Physics Department, State University of New York\\
Stony Brook, NY 11794, U.S.A.}\\
\ \\

\vspace{2cm}
{\large \bf Abstract}\\
\end{center}
We study the temperature dependence of quark antiquark correlations in the
space like direction. In particular, we predict the temperature dependence of
space like Bethe-Salpeter amplitudes using recent Lattice gauge data for the
space like string potential. We also investigate the effect of the space like
string potential on the screening mass and discuss possible corrections which
may arise when working with point sources.
\vspace{0.2cm}
\newpage
\noindent
\section{Introduction}
\noindent
It is generally believed that hadronic matter undergoes a transition to a new
phase, the so--called quark--gluon--plasma (QGP), which is a gas of weakly
interacting quarks and gluons.(For a general review see refs.
\cite{GPY81,Shu80}). As far as thermodynamic properties are concerned this
statement seems to be supported by lattice gauge calculations
(LGC)\footnote{For
a collection  of the most recent LGC results see ref. \cite{Lat92}.},
which show a
sharp rise of the energy density at the critical temperature $T_c$. Above
$T_c$, the energy-- and entropy density obtained in LGC are compatible with a
perturbative gas of quarks and gluons. Moreover
the quark condensate vanishes above
$T_c$ indicating the restoration of chiral symmetry. Furthermore, above $T_c$
screening masses extracted from spatial point to point correlation functions
assume the values of $M_{sc} = n \pi T$ \cite{DK87b,BGI91}, n being the
number of quarks involved. The $(\pi, \, \sigma)$ correlators give screening
masses smaller than the free value indicating considerable residual
interactions in these channels above $T_c$.
This is not too surprising as  these
modes
can be viewed as long range fluctuations in the order parameter of a second
order chiral restoration transition.

There exist, however, several phenomena which indicate a nontrivial
structure even above $T_c$. Simple thermodynamic
considerations \cite{KB93b}, for instance,
 show that about $50 \%$ of the zero temperature gluon condensate
still remains condensed above  the transition. Moreover,
while the large distance
behavior of space like point
to point correlators is consistent with free quarks, the measurement of
Bethe-Salpeter amplitudes in the space like direction shows very
strong correlations between quarks and antiquarks \cite{BDD91}. It has been
suggested \cite{HZ92} and demonstrated explicitly \cite{KSB92} that these
`observed' correlations can be related to the so--called space--like
string--tension,
which remains finite even above $T_c$ \cite{MP87}. This space like
string tension is extracted from the expectation value of a Wilson loop, which
loops only in the spatial direction. Contrary to the time like Wilson
loop, which measures the electric interactions and thus can be related to
the heavy quark potential, the space like Wilson loop is sensitive to magnetic
interactions. The associated string tension, therefore, does not relate
to a heavy quark potential. It rather provides a measure for  the strength
of a current-current type interaction, analogous  to that associated with
Amp\`{e}res law in electrodynamics. In contradistinction to the time like
string
tension, which vanishes above $T_c$, the space like string tension remains
finite at all temperatures \cite{Bor85,BFH93}

The effect of the spatial string tension on a quark antiquark pair
can be best understood if one transforms to so called `funny space', where
the euclidian time direction and one
of the spatial directions, say the `z'-direction,
are interchanged. At zero temperature, of course, real euclidian space
and funny
space are identical. At finite temperature, however, bosons
and fermions follow periodic and antiperiodic boundary conditions in the
temporal direction, respectively. In funny space these boundary conditions are
moved to the $z_f$-direction with no boundary conditions in the $\tau_f$
direction. Consequently a system at finite temperature in real space transforms
to one at zero temperature in funny space where quarks and gluons form
standing waves in the $z_f$-direction with momenta $p_q = (2n+1) \pi T$ and
$p_g = 2 n \pi T$, respectively, where n is an integer. Similarly, a spatial
Wilson loop $W_{x,z}$
in real space, which involves, say the $x$ and $z$ direction,
transforms into a time like Wilson loop $W^f_{x_f,\tau_f}$ in funny space.
Hence, it can be associated with a heavy quark potential in funny space.
The space-like Bethe-Salpeter amplitudes also obtain a `physical'
meaning in funny space.
They can be understood as wave functions of an quark-antiquark
pair.

First LGC with pure glue \cite{MP87} and more recent calculation involving
dynamical fermions \cite{GKH93} have
indicated that the space like string tension remains constant between zero
temperature (where it is identical with the time like string tension)
and temperatures slightly above the phase transition. In very recent
work by Bali et al. \cite{BFH93} have studied the temperature dependence of the
space like string tension in SU(2) gauge theory with high accuracy. For small
temperatures, $T \le 2 T_c$, they confirm that the string tension is
essentially independent of temperature. For
temperatures $T \ge 2 T_c$ they find that the string tension scales like
$\sigma(T) = (g(T) T^2)^2 \sigma_0$
where the T-dependence of the coupling $g(T)$
is given by the 2-loop $\beta$-function with a scale parameter $\Lambda = 0.076
T_c$. This form of the temperature dependence is what one would expect from
dimensional reduction arguments \cite{AP81} and the resulting value for the
string tension
agrees within $10 \%$ with that of three dimensional SU(2) gauge theory
\cite{Tep92}.

With the knowledge of the space like string tension as a function of
temperature, the resulting Bethe-Salpeter amplitudes can be calculated in the
model proposed in ref. \cite{KSB92}. This will provide a prediction for the
temperature dependence of these correlations which can be tested on the lattice
in order to check the validity of this model.
In this article we will present the wave functions for several temperatures. We
will also discuss several possibilities to better probe the detailed structure
of the quark-antiquark interaction and we will investigate the implications
for observables such as the screening masses etc.

This paper is organized as
follows: In the following section we briefly review the method we are going to
use to
calculate the Bethe-Salpeter amplitudes and present the resulting
amplitudes for different temperatures. Then we investigate other
possibilities to test the interaction more precisely.
In section three we discuss the
implications of the quark-antiquark interactions on the screening masses.
In the
final section we will compare our results with already existing LGC on the
temperature dependence of the wavefunctions \cite{SC93} and discuss possible
discrepancies.\\
\ \\
\section{Wavefunctions}
As mentioned in the introduction and explained in great detail in ref.
\cite{KSB92} at high temperature the
effect of the space like string tension on the Bethe-Salpeter amplitude is best
studied in `funny space'. There the
space like string tension acts just like a regular potential and the
boundary conditions, originally in the temporal direction, lead to standing
waves in the $z_f$ direction. At
high temperatures only the lowest momentum (Matsubara)
modes contribute, i.e. $p_z = \pm \pi T$ for the quarks/antiquarks, and the
momenta of
the standing waves $|p_z|$ simply act like an effective mass,
as far as the motion in the transverse direction ($x_f,y_f$) is
concerned.
Under these assumptions the space like
Bethe-Salpeter amplitude can be identified with the
wavefunction of a quark-antiquark pair with masses $m_q = \pi T$
in two dimensions (because the motion in the $z_f$ direction leads  to the
effective mass) interacting via the space like string potential.
At high
temperature, finally, the effective quark mass becomes large
so that a nonrelativistic approximation can be made.

Unless otherwise noted, from now on all variables and equations
are understood to be defined in
`funny space'. The subscript `f', indicating funny space variables, will,
therefore, be dropped.

The Schr\"odinger equation to be solved is \cite{KSB92}:
\be
- \frac{1}{\pi T}\left[\frac{1}{r} \frac{d}{dr}\left(r \frac{d }{dr}\right)
+ \frac{l^2}{r^2}  \right] \psi + V(r) \psi = E \psi
\label{eq.2.1}
\ee
where $l$ is the angular momentum reflecting the azimuthal symmetry and V(r)
is the potential which can be extracted from the space like Wilson loop.
Before we will show the actual wavefunctions let us first discuss some general
properties. For a stringlike potential $V(r) = -g/r + \sigma r + const$ the
large distance behavior of the wavefunction is governed by
\be
\left( -\frac{1}{\pi T} \frac{d^2}{d\,r^2}  + \sigma \,r \right) \, \psi = 0
\label{eq.2.2}
\ee
and consequently
\be
\psi \sim e^{\sqrt{\sigma \pi T} r^{3/2} } \,\,\,\,\,\,\, r \rightarrow \infty
\label{eq.2.3}
\ee
{}From this asymptotic form it is evident that the wavefunctions should
become
narrower with increasing temperature even if the string tension does not
increase! This is simply due to the increase of the effective quark mass $m_q =
\pi T$ in `funny space' or, equivalently, to the decrease of the de Broglie
wavelength of the quarks. An increase in the string tension would lead to an
even stronger fall off of the wavefunctions.

In fig. \ref{fig.2}a we show the ground state wave functions for
temperatures $T= .5,\ldots,8 T_c$ using the temperature
dependent potentials from ref.
\cite{BFH93}, which are plotted in fig. \ref{fig.1}. For comparison in fig.
\ref{fig.2}b the wavefunctions have been calculated with a potential fixed at
$T= 0.5 T_c$.
In both cases the wave functions become narrower with increasing
temperature, somewhat more in the first case where the string tension increases
with temperature as well. The difference between the wavefunctions in (a) and
(b) for a given temperature is not very large, however. As suggested by eq.
\raf{eq.2.3} a better way to illustrate the effect of the temperature dependent
potential is to plot the wavefunctions as  function of $r^{3/2} T^{1/2}$.
This is done in fig. \ref{fig.3}.
The wavefunction in the constant potential (fig. \ref{fig.3}b)
nicely follow the above scaling law, eq. \raf{eq.2.3},
(in case of $T=8 \,T_c$ one only can observe the
concave shape which would eventually lead to a slope parallel to the others.)
In fig. \ref{fig.3}a, on the other hand, the effect of the string potential is
clearly seen.
Although in principle the string tension can be read off from the asymptotic
slope, it requires the knowledge of the Bethe-Salpeter amplitudes over
several orders of magnitude and thus may not be feasible in practical LGC.
As can be seen from fig. \ref{fig.1}, over the range of the de Broglie
wavelength $d \sim 1/T$, where the bulk of the wavefunction is expected to be
located,
the effect of temperature on the potential is small. Thus,
in order to efficiently
probe the long range part of the interaction, one
has to find ways to move the wavefunction to larger distances.

One such possibility is to project on
states with finite orbital angular momentum ($l \neq 0$ in eq. \raf{eq.2.1}).
These states `feel' a centrifugal barrier $\sim l^2/r^2$ and, hence, are pushed
outside to larger distances as can be seen in fig.\ref{fig.4}, which shows the
wavefunctions for l=0,1,2,3 for a temperature $T = 2 \,T_c$. The
increased sensitivity on the potential is demonstrated in fig. \ref{fig.5}.
There the
ratio of the average radii for wavefunctions in the temperature dependent
and temperature independent potential ($<r (T-dep)> / <r(T-indep)>$)  with
\be
<r> = \int d^2r \,r \,|\psi(r)|^2
\label{eq.2.4}
\ee
is plotted as a function of the temperature for orbital
angular momentum $ l=0$
(open squares) and (l=3) full squares.  In case of $l=0$ the average
radius is
almost insensitive to the temperature dependence of the potential.
Even for the highest temperature, where  the string tension has increased
considerably, the difference in the mean radii of the wave functions is only
$\sim 10 \%$. Thus, due to the increase of the quark
effective mass, the ($l=0$) wavefunctions only `see'
 the short distance part of the potential, where the temperature
dependence is small (see fig. \raf{fig.1}). Only the tails of the wavefunctions
are sensitive to the linear rising part of the potential as demonstrated in
figs. \ref{fig.2} and \ref{fig.3}.
These of course affect the mean radius only very
little. In case of the $l =3$ wave functions, on the other hand, the
centrifugal potential $\Delta V  = l^2/r^2$ dominates at short distances and
pushes the wavefunctions out into the region where it becomes sensitive to the
string potential. At temperatures $T= T_c$ and $T= 1.33 \,T_c$, where the
string tension differs very little from the $T= 0.5 T_c$ value, an effect in
the ratio of the mean radii can already be observed. At high temperatures, the
`squeezing' of the state due to the increased string potential becomes very
clear\footnote{In principle the wavefunctions of the radially excited states
with $l=0$ also extend further out (see fig. \ref{fig.4}) and, hence, should be
more sensitive to the string potential. However, it seems  very
difficult to device sources and sinks which are sensitive to one particular
radial excitation only.}.

In summary, the bulk of
the s-wave ($l=0$) wave functions is mainly dominated by the effective quark
mass and information about the strength of the
string tension can be extracted only
from very precise LGC. The sensitivity can be increased, however, by projecting
on states with finite angular momentum. Thus sources and sinks in LGC should be
designed in a way that they produce states of
good angular momentum in the two
dimensional transverse (x,y) plane. Similar projections have already
been carried out
in the study of the glueball spectrum.
\ \\
\ \\
\noindent
\section{Screening masses}
In principle one would expect the temperature dependence of the space like
string potential to reflect on the screening masses. The eigenvalues
$E_{Schr}$ of the Schr\"odinger equation
are directly related to the screening
mass by
\be
m_{scr} = 2 \pi T + E_{Schr}
\ee
and an increase of the potential results in an increase of the eigenvalue.
But the eigenvalue is also affected
by the T-dependence of the effective quark mass,
because with increasing mass the wavefunction can sit deeper in the potential
well. This dropping of the eigenvalue as a result of the increasing quark mass
is demonstrated by the dashed line in fig. \raf{fig.6}, which gives
the eigenvalue as  a function of  temperature for the constant potential.
If the temperature dependence of the potential is taken into account as well
(full line) the eigenvalue does not change with temperature for $T \ge 2 \,
T_c$. Obviously the decrease of the eigenvalue due to the effective mass and
the increase as a result of the potential cancel each other with  a surprising
accuracy.

But more importantly,  the screening mass is dominated by the
`thermal' mass term $\sim 2 \pi T$. The energy eigenvalue $E_{Schr}$ is only a
small correction to this term\footnote{The nonrelativistic approximation,
which led to the Schr\"odinger equation \raf{eq.2.1}, implies, of course,
that the
mass term is much larger than the nonrelativistic energy eigenvalue.} and the
effect of the temperature dependence of the potential is even smaller. This is
demonstrated in fig. \ref{fig.6} (lower curves), where the ratio $M_{scr}/2 \pi
T$ is plotted. For all temperatures both temperature dependent and temperature
independent potential lead to the same screening mass. Consequently,
any effects from the T-dependence of the string potential on the
screening mass,
although possibly present, are not accessible on lattices of the present
size. For example for $T= 4 \, T_c$, $\Delta E_{Schr} \simeq 0.5 \, T_c$
and $M_{scr} \simeq 25 \,T_c$. Hence, in
order to be sensitive to the temperature dependence of the potential
a lattice which is about 200 times as
large in the spatial direction than in the time direction would be required !
\ \\
\ \\
\noindent
\section{Point sources}
Usually screening masses are defined as the logarithmic derivative of
point to point correlators taken at large distances. On the lattice, however,
the maximum separation between source and sink of the correlator is limited for
practical reasons. In this case the signal may be `contaminated' by an
admixture of excited states, because point source have an overlap with all
radial excitations of the quark antiquark-system and the limited separation
between sink and source may not be enough to screen the excited states
sufficiently. As we will show, in our case, the presence of excited states
hardly affects the results for the screening masses but leads to considerable
corrections for the Bethe-Salpeter Amplitudes.
Assuming the validity of the above potential model, the point to point
correlation function in funny space can be written as (for details see
appendix)
\be
C(\tau) = \frac{1}{V_4} \sum_{n} |\psi_n(0)|^2 e^{-M_n \tau}
\label{eq.4.1}
\ee
with $M_n = 2 \pi T + E_{Schr}$ and $V_4$ is the volume of the four-dimensional
euclidian box. Here
the contribution of states with higher Matsubara frequencies has been
neglected (see appendix). Because of the factor $|\psi(0)|^2$
only those states with angular momentum $l=0$ contribute to the sum in
eq. \raf{eq.4.1}. Similarly, for the Bethe-Salpeter amplitudes one has
(see appendix)
\be
\Psi(\tau,r_\perp) = \frac{1}{V_4} \sum_{n} \psi^*(0) \psi(r_\perp)
e^{-M_n \tau}
\label{eq.4.2}
\ee
As already mentioned, the ground  state is projected out by going to
large distances $\tau$. Usually this distance is assumed to be reached once
the screening mass
\be
M_{scr} = \frac{d}{d \tau} \ln( C(\tau))
\label{eq.4.3}
\ee
does not vary with $\tau$ anymore. We will demonstrate that this method may
have practical problems, if one wants to extract the ground state wavefunctions
as well. The argument is, that, due  to the dominance of the ` free, thermal'
contribution $\sim \pi T$ to the screening mass, the screening mass of the
ground state and of the first excited state differ only by a few percent.
Therefore, the screening mass appears to have assumed its asymptotic value even
if a nonnegligible admixture of excited states is still present. While these
excited states obviously do not alter the value of the screening mass,
they lead to
substantial corrections for the Bethe-Salpeter amplitudes. The reason is,
that at intermediate transverse distances $r_\perp$ the contributions to the
`wavefunction' of the
ground- and first excited state ($n=1,\: l=0$)
are comparable in magnitude but opposite in sign. Thus, the resulting
wavefunction appears narrower than that of the ground state\footnote{The
narrowing of the wavefunction due to the admixture of excited state is simply
a remnant of the initial $\delta$-function, which corresponds to a
sum over all
excited states, due to completeness. With increasing $\tau$ this
$\delta$-function disperses and eventually assumes the shape of the ground
state.}. This is demonstrated  in fig. \ref{fig.7}, where in part (a) we have
plotted the relative difference of the local screening mass $M_{scr}(\tau)$
(eq.
4.3) to that of the true ground state $M_{scr}(\infty)$ as a function of $\tau$
for temperatures $T = T_c$ and $T = 1.33
T_c$\footnote{We have selected these comparatively low temperatures, because
later on we want to discuss the results of \cite{SC93}, where the same
temperatures have been chosen. The  choice of temperature does not affect our
argument and we find the same results at higher temperatures.}. (For the
results presented here, the {\it four} lowest states (n=0,1,2,3) have been
taken into account.) At
$\tau = 1/T_c$ the resulting screening mass already differs by less than
$3 \, \%$ from that of the
ground state\footnote{Usually in LGC periodic boundary conditions are imposed
in the real-space spatial directions, which implies that only half the lattice
can be used to relax the wavefunction in funny space. Therefore, distances
$\tau$ much larger the $1/T_c$ are rarely realized.}. In (b) the same
ratio is plotted for the
average radii of the wavefunction as defined in eq. \raf{eq.2.4}.
At the same distance $\tau = 1/T_c$ the difference is still of the order of $20
\, \%$. This difference in the wavefunction is further illustrated in fig.
\ref{fig.8} where
we have plotted the radial density distributions $r_\perp |\psi(r_\perp)|^2$
for the true ground state wavefunction (thick lines) and for the
wavefunction as seen at a
distance $\tau = 1/T_c$. The narrowing of the wavefunction can clearly be seen.
This narrowing actually has been observed in the work of ref. \cite{SC93} where
the wavefunction as obtained from a LGC, using point sources, is compared with
the ground state wave function of the potential model for temperatures
$T= 0.66, \: 1.0, \: 1.33 \:T_c$.
The lattice used has a
spatial extent of exactly $2/T_c$, resulting in a maximum distance of
$\tau_f = T_c$
in funny space, because of periodicity. Thus, the conditions are identical
to those considered here.
For the highest temperature, where the
approximations of our potential model are best justified,
agreement could only be found if one assumed a quark mass which is twice as big
as given in the model\footnote{In ref. \cite{SC93} the calculation in the
potential model accidentally has been made using the full quark mass in the
Schr\"odinger equation instead of the reduced mass \cite{MC93}.
Thus, the agreement the
authors find for the highest temperature in effect means that one has
to double the quark mass.}. Although a fairly large current quark mass $m_q =
200 \, \rm MeV$ has been used, this cannot account for doubling the
effective quark mass of $\pi T$ at all.
The fact that a large effective quark mass is needed in order to reproduce the
lattice result implies, that the lattice wavefunction is narrower
than the one given by the potential model. This difference, of course, is
exactly what we would predict if an admixture of the excited states is still
present. For comparison in fig. \ref{fig.8} we have also plotted the ground
state wave function obtained after doubling the quark mass
(dashed-dotted line).
We find that it agrees fairly well with the effective wavefunction
extracted at a distance $\tau = 1/T_c$ (thin dashed line).

At this point we should note, that the narrowing of the apparent wavefunction
due to the superposition of excited states may also be relevant for the
measurement
of Bethe-Salpeter amplitudes at zero temperature.
Monitoring the screening mass in order to assure sufficient screening of the
excited states is only successful if the energy eigenvalue of the ground state
is smaller than the excitation energy of the first radial excitation (see
appendix for a more formal argument). This implies that the measurement of the
Bethe-Salpeter amplitudes for the pion
using this method should be fine while in
case of the $\rho$ the monitoring of the wavefunction itself is
advised.

The authors of ref. \cite{SC93} also find that the wavefunctions obtained
on the lattice for
the three different temperatures ($T= 0.66,\:1,\: 1.33T_c$)
are almost identical. In contrast our model would predict a narrowing of the
wavefunctions due to the increase of the effective quark mass $\sim \pi T$ (see
fig. \ref{fig.8}). (The changes in the
potential are minimal over the temperature interval considered here.)
Also for the more realistic
wavefunctions, where the presence of excited states due to point sources
is taken into account, the narrowing as a function of the temperature persists.
The ratio of the average radii between wavefunctions of different temperatures
is about the same for the true ground state wavefunction and that taken at a
distance $\tau = 1/T_c$ (see also fig. \ref{fig.7}b). If we
assume that the qualitative behavior of the space like string potential is the
same in SU(3) and in SU(2) gauge theory, the independence of the wavefunction
on the temperature as observed in ref. \cite{SC93}
cannot be understood within
our model. Of course the temperatures considered $T \le 1.33T_c$ are somewhat
low for the dimensional reductions arguments of our model to be good.
Furthermore, at
the lower two temperature points, $T=.66T_c$ and $T=T_c$, the
constituent quark mass may still be of considerable magnitude,
which also would affect the
wavefunction (the authors give a screening mass for the $\rho$ at $T=.66 T_c$
of about $\sim 5.7 T_c$ which should correspond to roughly
twice the constituent quark mass). The results of ref. \cite{SC93}, therefore,
may be at or below the lower limit of applicability of our model.
However, it would be very interesting to carry out LGC calculations at
larger temperatures $T \ge 2 T_c$ in order to see if the wavefunctions
eventually follow the prediction of our model or if the independence of the
temperature, as found in ref. \cite{SC93} persist even at temperatures where
one would expect dimensional reduction to be valid.

\section{Conclusions }
In this article we have studied the effect of the temperature dependence of the
space-like Wilson loop on the Bethe-Salpeter amplitudes in the space-like
direction. The temperature dependence of the wavefunctions is controlled not
only by the temperature dependence of the potential but also by the increase of
the effective quark mass with temperature in funny space ($m_q = \pi T$).
Because of that
`trivial' effect, only high accuracy LGC will be able to distinguish
between ground state wavefunctions from
temperature dependent and temperature independent string potential at
high temperatures. We, therefore, propose to project instead on states
with finite angular momentum in the transverse directions. As a result of the
centrifugal potential, these states are pushed towards larger transverse
distances and are thus much more sensitive to the string potential.

We have further
demonstrated, that  screening masses are extremely insensitive to the
temperature dependence of the potential.
They are essentially dominated by the `free thermal' value $\sim 2 \pi T$ and
deviations due to the string potential are very small.
We also have shown that the presence of excited states in the correlation
function due to point sources  minimally changes the
result for the screening mass. The Bethe-Salpeter amplitudes, on the other
hand,
come out much more narrow if excited states are still present. As a
consequence,
monitoring the screening mass in order to decide
if a state at a given distance has settled into the ground state
is not sufficient and
may lead to wrong conclusions about the actual size of the
Bethe-Salpeter amplitudes.

Finally, we have presented unique predictions for the temperature dependence of
the space like
Bethe-Salpeter amplitudes based on the potential extracted from high accuracy
SU(2) LGC.
In our model \cite{KSB92} these amplitudes are understood as Schr\"odinger
wavefunctions of quarks with mass $\pi T$ confined by the space like string
potential.
A measurement of these amplitudes on the lattice would, therefore,
provides an important test for the validity of our understanding of these
correlations.\\
\ \\
\noindent
{\bf Acknowledgements:}\\
I would like to thank F. Karsch for many interesting discussions. Most of
the research
for this work has been carried out at the ITP in Santa Barbara. For the nice
and fruitful atmosphere during the workshop on `Strong Interactions at Finite
Temperatures' I want to thank the staff of the ITP and the organizers of the
workshop, J. Kapusta and E. Shuryak. This work was supported in part by the
National Science Foundation under Grant No. PHY89-04035 and in part by the U.S
Dept. of Energy Grant No. DE-FG02-88ER40388.\\

\section{Appendix}

In this appendix we will derive how point to point correlation functions
and
Bethe-Salpeter amplitudes are expressed in terms of the eigenstates and
eigenvalues of the Schr\"odinger equation \raf{eq.2.1}. We will also give an
estimate on how the presence of excited states
affects the results for screening
masses and Bethe-Salpeter amplitudes. In the entire appendix we will work in
{\it funny space} i.e. we have periodic/antiperiodic boundary conditions in
z-direction and none in the $\tau$-direction. \\
\subsection{Point to point correlation functions}
The screening masses are determined from the exponential fall off in the $\tau$
direction of the point to point correlation functions. The correlation function
is given by:
\be
C(\tau,\vec{r}) = <0 |  j(\tau,\vec{r})\, j(0) |0>
\label{eq.a.1}
\ee
where $\tau \ge 0$ and $ r_3 = z \ge 0$ is assumed. Our main
assumption is that this correlation
function can be saturated by the eigenstates of the Schr\"odinger eq.
\raf{eq.2.1}. Therefore, we can insert a complete set of states
\be
1 = \sum_{n,p} \:\frac{|n,\vec{p}><n,\vec{p}|}{2 E_n(p)}
\ee
where $\vec{p}$ refers to the momentum of the center of mass motion and $n$ to
the quantum numbers of the internal motion, i.e. to that of the Schr\"odinger
equation. The correlation function can then be written as
\be
C(\tau,\vec{r}) = \sum_{n,p}\: \frac{1}{2 E_n(p)}
 <0 | j(\tau,\vec{r}) |n,\vec{p}><n,\vec{p}|j(0)  |0>
\label{eq.a.2}
\ee
Since  $|n,\vec{p}>$ are eigenstates of the momentum operator -- with discrete
eigenvalues for $\hat{p}_z$-- we can write
\be
 C(\tau,\vec{r}) = \sum_{n,p} \:\left|<0 | j(0)|n,\vec{p}> \right|^2
\exp( -E_n(p) \tau + i \vec{p}\vec{r})
\label{eq.a.3a}
\ee
By integrating over the spatial directions we project on states with zero
transverse momentum and  zero Matsubara frequency $p_z$ of the quark
antiquark pair.
\be
 C(\tau) = \sum_{n}\:\left|<0 | j(0)|n,\vec{p}=0> \right|^2
\exp( -E_n \tau )
\label{eq.a.3b}
\ee
The sum still includes all pairs of quark Matsubara frequencies
which add to zero, i.e. ($(2n+1)\pi T;\:-(2n+1)\pi T$). We will, however,
neglect the contribution of the states with $|n| >1$, because they have a
screening mass of $M_{scr} \ge 4 \pi T$ and, thus,
are screened very efficiently.

Finally the matrix element $<0|j(0)|n,\vec{p}>$ relates to the Schr\"odinger
wavefunction by
\be
<0 | j(0)|n,\vec{p}> = \sqrt{2 E_n(p)} \Psi_{n,p}(0)
\label{eq.a.4}
\ee
with
\be
\Psi(\vec{r}) = \frac{1}{\sqrt{\beta}} \Phi_{n,p}(\vec{r}) \psi_n(0)
\label{eq.a.5}
\ee
Here $\Phi(\vec{r})$ denotes the wavefunction of the center of mass
motion and
$\psi_n(0)$ is the wavefunction of eigenstates of the Schr\"odinger equation
describing the relative motion in the transverse $(x,y)$ direction. $\beta =
1/T$ gives the box size in the $z$-direction in funny space. The normalization
factor $1/\sqrt{\beta}$ is the remnant of the standing waves
$\sim \frac{\exp(-i p_z z)}{\sqrt{\beta}}$ for the motion in the
z-direction in the relative coordinate.
Therefore, with $|\Phi_n(0)|^2 = 1/V_3$ the correlation function is given by
($V_4 = V_3 \beta$)
\be
C(\tau) = \frac{1}{V_4}
\sum_{n}  \:  \left|\psi_n(0)\right| ^2
\exp( -E_n \tau)
\label{eq.a.6}
\ee
with $E_n = 2 \pi T + E_{Schr}$, where $E_{Schr}$ is the eigenvalue of the
Schr\"odinger equation \raf{eq.2.1}. As explained above, only the lowest
Matsubara frequency of the individual quarks are taken into account.

\subsection{Bethe-Salpeter amplitudes}
The Bethe-Salpeter amplitude is defined as
\be
B(\tau,\vec{r},r_\perp) = <0|\left( \bar{\psi}(\tau,r-\frac{r_\perp}{2})
\Gamma \psi(\tau,r+\frac{r_\perp}{2})\right) \:\: j(0) | 0>
\label{eq.a.7}
\ee
where $\Gamma$ denotes the appropriate spin-isospin matrix, which we will, for
simplicity, ignore from now on.
Again we can write the product
\be
<n,\vec{p}|\bar{\psi}(\tau,r-\frac{r_\perp}{2})
\Gamma \psi(\tau,r+\frac{r_\perp}{2})  |0>
= \frac{\sqrt{2 E_n(p)}}{\sqrt{\beta}} \Phi_{n,p}(\vec{r}) \psi(r_\perp)
\label{eq.a.8}
\ee
(Note, that $r_\perp$ is in the transverse $(x,y)$ direction only.)
By inserting a complete set of states, we can factor out the c.m. motion and
obtain
\be
B(\tau,\vec{r},r_\perp) = \sum_{n,p} <n, \vec{p}| j(0)| 0>
\frac{\exp(-E_n \tau +
i \vec{p} \vec{r})}{\sqrt{2 E_n}} \psi_n(r_\perp)
\label{eq.a.9}
\ee
and
after integrating over the spatial coordinates the states $|n,\vec{p}=0>$ with
vanishing c.m. momentum are projected out. Using eq. \raf{eq.a.4} we finally
have
\be
B(\tau,r_\perp) = \frac{1}{V_4} \sum_{n} \psi^*_n(0) \psi_n(r_\perp)
\exp(-E_n \tau)
\label{eq.a.10}
\ee

\subsection{Effect of excited states on screening mass and Bethe-Salpeter
amplitudes}
Finally let us give an estimate on how the presence of excited states affects
the screening mass and the Bethe-Salpeter amplitude. Let us, for simplicity,
consider only 2-states. Then the correlation function is given by (up to a
constant)
\be
C(\tau) = \exp(-E_0 \tau) + \alpha^2 \exp(-E_1 \tau)
\label{eq.a.11}
\ee
with
\be
\alpha^2 = \left| \frac{\psi_0 (0)}{\psi_1(0)} \right|^2
\label{eq.a.12}
\ee
The screening mass at given distance $\tau$ follows from the logarithmic
derivative of the correlation function
\be
M_{scr} = - \frac{d}{d \tau} \, \ln C(\tau) = E_0\, (1 + \Delta)
\label{eq.a.13}
\ee
with
\be
\Delta = \alpha^2 \frac{E_1-E_0}{E_0} \exp(-(E_1-E_0) \tau)
\label{eq.a.14}
\ee
The screening mass is, thus, unaffected by the presence of the excited state if
$|\Delta | \ll 1$. This is achieved by going to a sufficiently large distance
$\tau$.

For the Bethe-Salpeter amplitude or wavefunction, on the other hand, one has
\be
\psi(r_\perp)& =& \psi_0(r_\perp) \exp(-E_0 \tau ) + \alpha \psi_1(r_\perp)
\exp(-E_1 \tau)
\nonumber \\
\alpha & = &\pm \left| \frac{\psi_0(0)}{\psi_1(0)} \right|
\label{eq.a.15}
\ee
which can be related to the accuracy of the screening mass $\Delta$
\be
\psi(r_\perp) = \psi_0(r_\perp) \exp(-E_0 \tau ) \left( 1 \pm
\frac{\Delta}{\alpha^2} \frac{E_0}{E_1 - E_0} \right)
\label{eq.a.16}
\ee
Typically $\alpha^2 \le 1$ (see e.g. fig. \ref{fig.4}) and hence
the accuracy of the Bethe-Salpeter amplitude $\Delta_{B.S.}$ is
\be
\Delta_{B.S.} \ge \Delta_{S.M.} \frac{E_0}{E_1 - E_0}
\label{eq.a.17}
\ee
Therefore, the error in the wavefunction may be considerably larger than that
in the screening mass if the energy eigenvalue of the ground state is much
larger than the excitation energy of the first excited state, i.e. $E_0 \gg E_1
- E_0$.

\newpage

{\bf Figure captions}
\begin{figure}[h]
\caption{Space like string potential from ref.\protect\cite{BFH93}}
\label{fig.1}
\end{figure}

\begin{figure}[h]
\caption{Space like Bethe-Salpeter amplitude for ground state for temperature
dependent potential (a) and for potential fixed at $T=0.5 \, T_c$ (b).}
\label{fig.2}
\end{figure}

\begin{figure}[h]
\caption{Space like Bethe-Salpeter amplitude for ground
state for temperature
dependent potential (a) and for potential fixed at $T=0.5 \, T_c$ (b).
Here x-axis has been rescaled to demonstrate the asymptotic behavior.}
\label{fig.3}
\end{figure}

\begin{figure}[h]
\caption{Space like Bethe-Salpeter amplitude for ground state (n=0,l=0), the
first radial excitation (n=1,l=0),
and for finite angular momentum (n=0,l=1,2,3).
The temperature is $T = 2 \, T_c$.
}
\label{fig.4}
\end{figure}

\begin{figure}[h]
\caption{Ratio of average radii between temperature dependent and temperature
independent potential for states with l=0 (full line) and l=3 (dashed line)}
\label{fig.5}
\end{figure}

\begin{figure}[h]
\caption{Eigenvalues of the ground state
of the Schr\"odinger equation and associated screening masses
for temperature dependent (solid line) and temperature independent (dashed
line). The dotted line gives the screening mass of the first radial
excitation.}
\label{fig.6}
\end{figure}

\begin{figure}[h]
\caption{Apparent screening mass (a) and average radii of the apparent
wavefunctions as a function of separation of source and sink $\tau$}
\label{fig.7}
\end{figure}

\begin{figure}[h]
\caption{Probability density distribution for the true
ground state (thin lines) and for the effective wavefunction as obtained at a
distance of $\tau = 1/T_c$ (thick lines).
The dashed-dotted line corresponds to the
wavefunction obtained with twice the quark mass for $T = 1.33\,T_c$}
\label{fig.8}
\end{figure}

\end{document}